\documentclass[conference]{IEEEtran}
\usepackage{cite}
\usepackage[usenames,dvipsnames]{xcolor}
\usepackage[pdftex]{graphicx}
\usepackage{circuitikz}
\usepackage[hidelinks,bookmarks=false]{hyperref}
\usepackage{booktabs,colortbl}
\usepackage{multirow}
\usepackage[cmex10]{amsmath}
\usepackage{mathrsfs}
\usepackage{bbold}
\usepackage{array}
\usepackage{multirow}
\usepackage{booktabs,colortbl}
\usepackage{amssymb}
\usepackage{siunitx}
\usepackage{amssymb}
\usepackage{cases}
\usepackage{algorithm}
\usepackage{algpseudocode}
\usepackage{color,soul}
\usepackage{tikz}
\usetikzlibrary{decorations.pathreplacing}
\usetikzlibrary{arrows,shapes,positioning}
\usetikzlibrary{patterns}
\usepackage{pgfplots}
\pgfplotsset{compat=newest}
\pgfplotsset{plot coordinates/math parser=false}
\newlength\figureheight
\newlength\matlabfigurewidth
\usepackage{multicol}
\usepackage{breqn}
\usepackage{cuted}
\usepackage{amsthm}

\theoremstyle{definition}
\newtheorem{definition}{Definition}[section]
\theoremstyle{remark}

\usepackage{mathtools}
\newcommand{\argmin}[1]{\underset{#1}{\text{argmin}}\,}

\newcommand\numberthis{\addtocounter{equation}{1}\tag{\theequation}}

\ifCLASSINFOpdf
\else
\fi
\usepackage{amsmath}
\begin{document}
%
\title{Constrained hierarchical networked\\optimization for energy markets}

\author{\IEEEauthorblockN{Lorenzo Nespoli}
\IEEEauthorblockA{Swiss Federal Institute of Technology\\Lausanne, Switzerland\\
	University of Applied Sciences and\\ Arts of Southern Switzerland\\Lugano, Switzerland\\
Email: lorenzo.nespoli@epfl.ch}
\and
\IEEEauthorblockN{Vasco Medici}
\IEEEauthorblockA{University of Applied Sciences and\\ Arts of Southern Switzerland\\Lugano, Switzerland\\
Email: vasco.medici@supsi.ch}}


%


\maketitle

\begin{abstract}
In this paper, we propose a distributed control strategy for the design of an energy market. The method relies on a hierarchical structure of aggregators for the coordination of prosumers (agents which can produce and consume energy). The hierarchy reflects the voltage level separations of the electrical grid and allows aggregating prosumers in pools, while taking into account the grid operational constraints. 
To reach optimal coordination, the prosumers communicate their forecasted power profile to the upper level of the hierarchy. Each time the information crosses upwards a level of the hierarchy, it is first aggregated, both to strongly reduce the data flow and to preserve the privacy.
In the first part of the paper, the decomposition algorithm, which is based on the alternating direction method of multipliers (ADMM), is presented. In the second part, we explore how the proposed algorithm scales with increasing number of prosumers and hierarchical levels, through extensive simulations based on randomly generated scenarios.
\end{abstract}


%
\IEEEpeerreviewmaketitle

\section{Introduction}
With the penetration of renewable energy sources (RES), a decentralized market design with self-dispatch components is developing in the distribution grid. The demand side is becoming increasingly capable of providing flexibility services and contributing to a reliable power system and price stability on power markets.
As flexible generation and consumption capacity will be highly fragmented and distributed, to better exploit it and maximize its economical profitability a high number of prosumers will be required to coordinate with each other, when responding to demand response (DR) signals. A market design that rewards flexibility needs to be set up. The highly stochastic nature of RES generation calls for a market that is able to operate in near real-time.\\

The presence of highly correlated distributed generation increases the risk of local congestions and voltage fluctuations. Indeed, the highly stochastic nature of RES generation calls for near real-time control of flexibility.
Many authors have tried to achieve prosumers coordination in different ways, for instance by modelling the market as Cournot games with constraints \cite{Yao2007},\cite{Yang2010},\cite{Kian2005}, seeking Nash equilibria in non-cooperative games \cite{Kim2013a},\cite{Park2016}, generalized Nash equilibria \cite{Li}, and as a distributed control problem.
An optimal coordination of the prosumers can be achieved in different ways, among which making use of aggregators is one of the most promising ones \cite{Parag2016}. The question of how the aggregators will achieve coordination, however, is still matter for research with a number of promising solutions being investigated.
An interesting way to exploit flexibility of a pool of prosumers is to explicitly formulate a common target for their aggregated power profile, and give them economic incentives to follow this target. Energy retailers and balance responsible parties, which bid for purchase of energy in the energy market, would benefit from a reduction of uncertainty in the prosumers consumption or production.
In this paper, we consider prosumers as cooperative agents not able to modify the control algorithm that optimizes the operations of their flexible loads. As such, we are not obliged to choose prosumers' utility functions that generate a unique generalized Nash equilibrium. We will rather focus on a distributed control protocol allowing prosumers’ coordination through multiple voltage levels.
In this context, a good coordination protocol must ensure prosumers’ privacy while being scalable. Prosumers’ privacy is inherently guaranteed if they do not need to share their private information (e.g. size of batteries, desired set-point temperatures in their homes), or their forecasted power profile. Scalability ensures that the computational time of coordination scales near-linearly with increasing number of agents, allowing for fast control.\\
Most studies on the subject are focused on maximizing the welfare of a group of prosumers, by means of maximizing their utility functions. In the mathematical optimization framework, this problem can be modelled as an allocation or exchange problem \cite{Boyd2010}.
In \cite{HomChaudhuri2014} the welfare maximization problem is considered with additional coupling constraints, modelling line congestions. The problem is solved using a primal-dual interior point method, considering that each agent has access to the dual updates of his neighbours. In \cite{Ghadimi2013} the same problem is solved with different multi-steps gradient methods. 
In recent years, other authors proposed decomposition techniques based on the theory of monotone operator splitting\cite{Bauschke2011}. These algorithms are known to have more convenient features in terms of convergence with respect of the gradient-based counterparts. An example of such approach, is represented by the proximal algorithms, which are well suited for non-smooth, constrained, distributed optimization \cite{Parikh2013}.
In \cite{Margellos2017} the decomposition of the welfare maximization problem under uncertainty is considered, combining proximal gradient method and weighted gradient method. \cite{Moret} proposes a decentralized energy market that makes use of the alternate direction method of multipliers (ADMM) \cite{Boyd2010} to split the problem. In \cite{Jian2006} the unit commitment problem is solved through ADMM. In \cite{Diekerhof2017} a robust implementation of demand response mechanism is introduced and solved with ADMM.
A multi-objective optimization problem aiming at maximizing prosumer's welfare while minimizing a system-level objective can be modelled as a sharing problem, see for example \cite{Hong2015,Boyd2010}. The general sharing problem can be written as:
\begin{equation} \label{eq:monolithic}
\argmin{x \in \mathcal{X}} e(x) + \sum_{i=1}^N f_i(x_i) 
\end{equation}
where $x_i \in \rm I\!R^t$ are the prosumers' vector of decision variables, $x = [x_i ^T]^T = [x_1^T,...x_N^T]^T \in \rm I\!R^{NT}$ is the concatenation of all the decision variables, $\mathcal{X}$ is a convex and compact set of constraints, $e: \rm I\!R^{NT} \rightarrow \rm I\!R$ is a system-level objective function and $f_i: \rm I\!R^T \rightarrow \rm I\!R$ is a prosumer specific objective function.
For example, in \cite{Halvgaard2014c} an application to the energy market is considered, where a dynamic version of the sharing problem with a tracking profile system-level objective is decomposed using the Douglas-Rachford splitting. 
The same problem is solved in \cite{Braun}, where an adaptation of the ADMM algorithm for the sharing problem is used. This is the same solution approach proposed in \cite{Boyd2010}, \S 7.3.
In \cite{Bhattarai2013} a high-level hierarchical control flow between the DSO, independent aggregators and prosumers is proposed, but no link is given between the different control signals. It is worth noting that some authors, as \cite{Diekerhof2017}, \cite{Braun},\cite{Jian2006}, use the term hierarchical to refer to a single level hierarchy, in which a the problem can be solve with a master-slave solution scheme.
\\
In this paper, we introduce a hierarchical market design that exploits the flexibility of prosumers located in different voltage levels of the distribution grid. By aggregating a higher number of prosumers, we can better exploit their flexibility for grid regulation. Each prosumer communicates with an aggregator, i.e. with his parent node.
The algorithm, which can be monolithically described as a sharing problem, effectively preserves privacy between the different levels, since only aggregated information is available at the higher levels of the communication structure. In the first part of the paper, we present the algorithm used to solve the coordination problem. In the second part, we present results from the coordination of prosumers in different hierarchical structures. We systematically vary the number of levels and draw the number of prosumers per level from a uniform distribution. Results on convergence and computational time are presented.

\section{Problem formulation}
In this work we jointly maximize prosumers' specific objective functions  and a system-level objective, taking into account grid constraints. Prosumer's flexibility is modeled by means of electrical batteries, but the approach can be generalized to other kinds of flexibilities. Prosumers communicate only indirectly, with the help of aggregators, located in the branching nodes of the hierarchical structure.


This problem can be formulated monolithically using the very general formulation in \ref{eq:monolithic}. However, \ref{eq:monolithic} does not explicitly show the tree-like dependences of the problem we would like to solve. 
In order to express the hierarchical nature of the problem, we briefly introduce the nomenclature of rooted tree structure, from graph theory. A rooted tree $\tau$ is a unidirected acyclic graph, with every node having exactly one parent, except for the root node. Each node is identified by a tuple $(d_1..d_l...d_{l_d})$ where $l_d$ is the level to which the node belongs, and each entry represent the enumeration of its $l_{th}$ level ancestor. In this paper, we keep $d_1 = 1$, and indicate the root node as $\emptyset$. 
Next, we introduce the definition of the set we will use in the description of the algorithm.
\begin{definition}[Node sets]
	\begin{center}
		\begin{enumerate}
			\item Descendants of node $ A =(d_1,..d_{L})$. $\mathcal{D}(A) = \big\{ A_j = (j_1,..j_{L_j}): L_j>L, (j_1,..,j_{L}) = A \big\} $
			\item Leaf node $\mathcal{L}(\tau) = \big\{A \in \tau: \mathcal{D}(A) = \emptyset \big\}$ 
			\item Nodes in level $l$. $\mathcal{N}_l(\tau) = \big\{A\in \tau : L = l\big\}$
			\item Branching nodes. $\mathcal{B}(\tau) = \setminus \mathcal{L}(\tau)$ 
			\item Anchestors of node $A$. $\mathcal{A}(A) = \big\{ A_j\in \tau : A \in  \mathcal{D}(A_j)\big\}$
		\end{enumerate}	
	\end{center}
\end{definition}
\begin{figure}[h]
	\centering
	\includegraphics[width=0.4\linewidth]{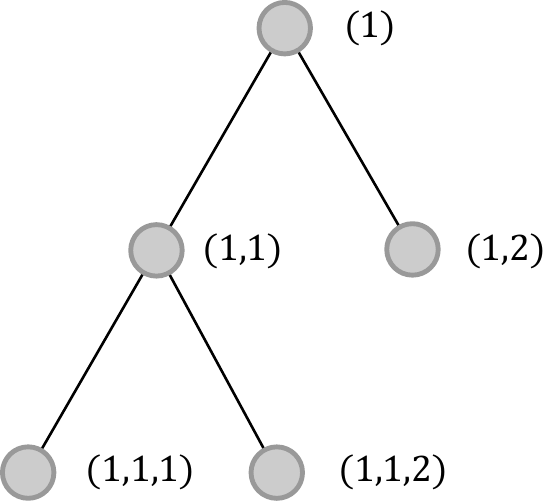}
	\caption{Example of rooted tree hierarchical structure}
	\label{fig:hierstr}
\end{figure}
Figure \ref{fig:hierstr} shows a simple example of rooted tree hierarchical structure and the tuples associated with every node.
Now, we can use the definition of branching node of level $l$ to rewrite problem \ref{eq:monolithic}. For sake of simplicity and clarity of exposition, we assume to have only one constraint in each branching node. We will remove this assumption in the simulations. 
\begin{equation}\label{eq:the_problem}
\begin{aligned}
\argmin{x} & e(S_{\emptyset}x) + \sum_{i=1}^N f_{c,i}(x_i) \\
s.t.:\ & S_B x  \leq v_B \quad \forall \ B \in \mathcal{B} (\tau)\\
\end{aligned}
\end{equation}
where $x_i$ are the actions associated to the agent $i$ in the terminal node $A_i$, $B$ denotes the branching nodes of the tree $\tau$ and $S \in \rm I\!R^{T \times N} $ are summation matrices defined as:
\[
S_B = [M_{B,A_j}], \quad  
M_{B,A_j} = \begin{cases}
a_{B,j} \rm I\!I_T,              & \text{if} \quad A_j  \in  \mathcal{D}(B)\\
\mathbf{0}_T, & \text{otherwise}

\end{cases}
\]
where $\mathbf{0}_T$ and $\rm I\!I_T$ are the zero and identity matrix of size $T$ respectively, $a_{B,j}$ is a weight associated to the $j_{th}$ descendant of branching node $B$ and $f_{c,i}$ is the objective function of agent $i$, which takes into account his agent-specific constraints:
\[
f_{c,i}(x_i) = 
\begin{cases}
f(x_i), & \text{if } \quad x_i \ \in \ \mathcal{X}_i\\
\infty,              & \text{otherwise}
\end{cases}
\]
We can now clarify how the notation used in \ref{eq:the_problem} allows us to express grid constraints in a flexible way. Again, for sake of simplicity, in the following we only consider apparent power and no additional uncontrolled loads. Both these assumptions will be removed in the presented simulations.
Since we are considering a radial grid, we can express the total power at a given branching node $B$ as the sum of the power of its descendants, $\mathcal{D}(B)$. In this case, we can set $a_{B,j} = 1 \quad \forall j $ so that $S_B$ becomes the time-summation matrix of powers of $\mathcal{D}(B)$.
Imposing voltage constraints in terms of apparent power involves solving the power flow (PF) equations.
Solving the exact PF equations would result in a non-convex optimization problem, which are in general difficult to solve. Different convex formulations of the PF exist, the most adopted being the DC PF model. Despite well suited for high and medium voltage grids, this model is typically inappropriate for distribution systems \cite{Molzahn2017}. Furthermore, the DC PF still requires susceptances, voltage angles and the knowledge of the grid topology. Differently from the medium voltage grid, parameters and topology are hardly available for low voltage grids. A better linear approximation for low-voltage grids is represented by the first order truncation of the PF equations \cite{Almasalma2017}:
\begin{equation}\label{eq:voltage}
\vert V \vert \approx V_0 + P^T\nabla_P \vert V \vert   + Q^T \nabla_Q \vert V \vert 
\end{equation}
where $P \in \rm I\!R^n $, $Q \in \rm I\!R^n$  are the nodal active and reactive power in a grid of $n$ nodes and $V$ and $V_0$ are the voltage and reference voltage at a given point of the grid. $\nabla_P \vert V \vert \in \rm I\!R^n $ and $\nabla_Q \vert V \vert \in \rm I\!R^n $ are the gradients with respect to the nodal active and reactive power at each node, and are collectively called voltage sensitivity coefficients.
It has been shown that they can be estimated using distributed sensor networks of phasor measurement units \cite{Mugnier2016} or even smart meter data \cite{Weckx2015}. 
We can use this approximation, replacing $a_{B,j}$ with the voltage sensitivity coefficient of node $j$ with respect to $B$. If we set $v_B = V_{max,B} -V_{0,B}$ we retrieve the formulation in \ref{eq:voltage}.

\section{Problem decomposition}
A trivial way of decomposing the problem would consist in repeatedly applying existent decomposed formulation of the sharing problem for each level. This would result in an exponentially increasing computational time, with the number of considered levels, namely:
\begin{equation}
N_{tot} \sim t n N_i ^L  \prod_{l = 1}^{L} N_b ^l \sim    N_i ^L N_b^{\frac{L+L^2}{2}}
\end{equation} 
where $t$ is the computational time for each agent for solving its local problem, n is the number of agents per level, $N_i$ is the number of iterations before convergence for a single level, $L$ is the number of levels,  and $N_b$ is the number of branches per level. Instead of following this strategy, we decompose the monolithic formulation of the problem to obtain a near-linear increase of iterations with the number of levels.
Problem \ref{eq:the_problem} is not decomposable as it is, due to the first term $e(x)$ and the coupling constraints. To decompose it, we introduce additional variables, copy of the linear transformations of $\mathcal{D}\mathcal{B}(\tau)$: 

\begin{equation}\label{eq:the_problem_y}
\begin{aligned}
\argmin{x} & e(y_{\emptyset}) + \sum_{i=1}^N f_{c,i}(x_i) \\
s.t.:\ & y_{B,i}  = a_{B,i} x_i \quad \forall \ B \in \mathcal{B} (\tau), \quad \forall i \in \mathcal{D}_{\mathcal{L}}(B) \\
& \sum_{i \in \mathcal{D}_{\mathcal{L}}(B)} y_{B,i} \leq v_B \quad \forall \ B \in \mathcal{B} (\tau)\\
\end{aligned}
\end{equation}
where $\mathcal{D}_{\mathcal{L}}(B)$ is the set containing the terminal nodes which descent from branch $B$, $\mathcal{D}_{\mathcal{L}}(B) = \mathcal{D}(B) \cap \mathcal{L}(\tau)$
We now formulate \ref{eq:the_problem_y} as an unconstrained minimization problem. We do it using an augmented Lagrangian formulation for the equality constraints involving duplicated variables and explicitly consider inequality constraints of \ref{eq:the_problem_y} by means of the indicator functions. See for example \cite{Parikh2013} \S 5.4 for a similar approach applied to the allocation problem. 

\begin{multline}\label{eq:lagrangian}
\mathcal{L}_{\rho} = e(y_\emptyset) + \sum_{i=1}^N f_{c,i}(x_i) +I_{\mathcal{Y}_{B}}(\overline{y}_{B}	 )\\
+\sum_{B\in \mathcal{B} (\tau)}\sum_{i \in \mathcal{D}_{\mathcal{L}}(B)} \frac{1}{2\rho} \Vert a_{B,i} x_i - y_{B,i}  +\lambda_{B,i} \Vert_2^2 
\end{multline}

where $\overline{y}_{B}$ is the sum of $y_{B,i}$, defined as $\overline{y}_{B} = \sum_{i \in \mathcal{D}_{\mathcal{L}}(B)} y_{B,i}$,  $\lambda_{B,i}$ are the dual variables associated to the equality constraints of problem \ref{eq:the_problem_y}, $\rho \in \rm I\!R$ is the augmented Lagrangian parameter, $\mathcal{Y_B}$ are the constraint sets of the inequalities of problem \ref{eq:the_problem_y} and $I_{\mathcal{Y}_B}(y_B)$ are indicator functions, defined as:
\[
I_{\mathcal{Y}}(y) = 
\begin{cases}
0, & \text{if} \quad y \ \in \ \mathcal{Y}\\
\infty,              & \text{otherwise}
\end{cases}
\]
and $y_B = [y_{B,i}]$.
We now follow the alternating direction method of multipliers (ADMM) strategy \cite{Boyd2010}, and perform a joint minimization-maximization of $\mathcal{L}_{\rho}(x,y,\lambda)$. Note that the convergence results from the ADMM algorithm allow us to take into account extended-real-valued and non-differentiable functions, as the indicator function. The overall decomposed problem can be written as:

\begin{align*}
x_i^{k+1} &= \argmin{x_i} f_{c,i}(x_i) +\\
&  \sum_{A\in \mathcal{A}(A_i)}  \sum_{j \in \mathcal{D}_{\mathcal{L}}(A)} \frac{1}{2\rho} \Vert a_{A,j} x_j^k -y_{A,j}^k  +\lambda_{A,j}^k \Vert_2^2 \numberthis \\
y_{\emptyset}^{k+1} &= \argmin{y_{\emptyset}} e(y_{\emptyset}) +\\
&\sum_{i \in \mathcal{D}_{\mathcal{L}}(\emptyset)} \frac{1}{2\rho} \Vert a_{\emptyset,i} x_i^{k+1} - y_{\emptyset,i}  +\lambda_{\emptyset,i}^k \Vert_2^2 + I_{\mathcal{Y}_{\emptyset}}(\overline{y_{\emptyset}} ) \numberthis \\
y_{B,i}^{k+1} &= \argmin{y} \frac{1}{2\rho} \Vert a_{B,i} x_i^{k+1} -y  +\lambda_{B,i}^k \Vert_2^2 + I_{\mathcal{Y}_B}(\overline{y}_{B}) \numberthis \\
\lambda_{B,i}^{k+1} &= \lambda_{B,i}^k +\rho (a_{B,i} x^{k+1} -y_{B,i}^{k+1} ) \numberthis
\end{align*}

Note that due to the definition of the $S_B$ matrices, the $x_i$ update only involves constraints from the ancestors of node $A_i$. This is thanks to the fact that in our model, actions of node $A_i$ do not influence agents in other subtrees, and is the ultimate reason that justifies a hierarchical communication structure. 
Note that for solving the first minimization problem, agent $i$ must consider all the other agent actions $x_{-i}$ as fixed. It has been shown that following a Gauss-Seidel like iteration, in which agents update their actions in sequence, considering all the available updated actions from the other agents, the above formulation converges \cite{Hong2015}. Anyway, this requires to solve the subproblems in sequence, reducing the computational advantage of a distributed solution. We prefer to use a parallel formulation, in which agents solve their own problems simultaneously. This will obviously not decrease the overall computations, but rather the effective convergence time.
We parallelize the problem fixing the average of the auxiliary variables $y_{B,i}$ during their update step. This will effectively reduce the overall number of variables and allows for a stable parallelization. Note that the resulting formulation can be interpreted as requiring each $x_i$ to reduce the average constraint violations $S_Bx^k -\overline{y}_{B}$ minus the scaled value of the previous iteration $a_{B,i}x^k$. See \cite{Braun} and \cite{Boyd2010} \S 7.3 for a detailed description of the method. We reformulate the iterations noting that the $\overline{y}_B$ updates can be rewritten in terms of the proximal operator $\mathbf{prox}_{\rho f}$. Additionally, for all the $\overline{y}_B$ but the root node, the proximal operator of the indicator function reduces to the projection operator $\boldsymbol{\Pi}_{\mathcal{X}} $.

\begin{align*}
x_i^{k+1} &= \argmin{x_i} f_{c,i}(x_i) +\sum_{B\in \mathcal{A}(A)}  \frac{1}{2\rho} \Vert (S_B x^k -\overline{y}_{B}^k)/N_{B}\\
&-a_{B,i}x_i^k +x_i +\overline{\lambda}_B^k\Vert_2^2 \label{eq:agent_min} \numberthis \\
\overline{y}_{\emptyset}^{k+1} &= \boldsymbol{\Pi}_{\mathcal{Y}_{\emptyset}}(\mathbf{prox}_{\rho e} (S_{\emptyset} x^{k+1}  +\overline{\lambda}_{\emptyset}^k )) \numberthis \label{eq:root_min} \\
\overline{y}_B^{k+1} &= \boldsymbol{\Pi}_{\mathcal{Y}_B} (S_B x^{k+1} +\overline{\lambda}_B^k) \numberthis \\
\overline{\lambda}_B^{k+1} &= \overline{\lambda}_B^{k} +\frac{\rho}{N_{B}} (S_B x^{k+1} -\overline{y}_B^{k+1})\label{eq:dual_max} \numberthis 
\end{align*}

where $N_B$ is the number of descendants of branch $B$. Since the root node update involves the minimization of system-level objective function $e$, equation \ref{eq:root_min} projects its proximal minimization into the root node constraint set $\mathcal{Y}_{\emptyset}$, similarly to proximal gradient methods, as the forward-backward splitting \cite{Stella2017a}.       

The pseudocode of the update rule is summarized in Algorithm \ref{alg:1}. The sum of norms in the agent update step can be reduced to a single norm:
\begin{equation}
x_i^{k+1} = \argmin{x_i} f_{c,i}(x_i) +\frac{1}{2\rho} \Vert r_i + R_i x_i- R_{a,i}x_i^k\Vert_2^2
\end{equation}  
where $R_i = [\rm I\!I_T] \in \rm I\!R^{Tn_a \times T} $ is the concatenation of $n_a$ identity matrices where $n_a$ is the number of ancestors of agent $i$,  $R_{a,i} = [a_{B,i}\rm I\!I_T] \in \rm I\!R^{Tn_a \times T} $,  and $r_i = [r_B] \in \rm I\!R^{Tn_a}$ is the  concatenation of reference signals form its ancestors:
\begin{equation}
r_B = (S_B x -\overline{y}_B)/N_B+\lambda_B
\end{equation}

\begin{algorithm}
	\caption{Hierarchical optimization}\label{alg:1}
	\begin{algorithmic}[1]
		\State Initialize $\mathrm{err}=tol*2,\overline{y}_B = 0_T, \lambda_B = 0_T$ 
		\While{$\mathrm{err} \leq tol$} 
		\State $x_i^{k+1} \leftarrow x_i^k,r_i^k  $ \Comment{agents}
		\State $\overline{y}_{A}^{k+1}, r_B^{k+1} \leftarrow \overline{\lambda}_{B} ,x_{\mathcal{D}(B)}^{k+1}$\Comment{branch $B$}
		\State $\mathrm{err}_B^{k+1} \leftarrow x_{\mathcal{D}(B)}^{k+1},\overline{y}_{B}^{k+1}$\Comment{primal err in branch $B$}
		\State $\overline{\lambda}_{B}^{k+1} \leftarrow \overline{\lambda}_{B}^{k}, \mathrm{err}^{k+1}$ \Comment{dual variable in branch $B$}
		\State $err \leftarrow err_B $
		\EndWhile
		
	\end{algorithmic}
\end{algorithm}

We can see from the pseudocode in \ref{alg:1} that each agent requires only the reference signals from its ancestors to solve its optimization problem.
Thanks to the hierarchical communication structure, these signal can be collected from the parent node of agent $i$. 
This allows the algorithm to be solved in a forward-backward passage. In the forward passage each branch $B$ sends its reference signal $r_B$ and the one received by its parent to his children, which propagate it downwards through the hierarchy. At the same time, prosumers in leaf nodes solve their optimization problem as soon as they receive their overall reference signal $r_i$. In the backward passage agents send their solutions to their parents, which collect them and send the aggregated solution upward.
Note that $r_B$ contains only aggregated information from branch $B$, which ensures privacy among prosumers.

\section{Simulation results}
\begin{figure*}[h]
	\centering
	\includegraphics[width=0.9\linewidth]{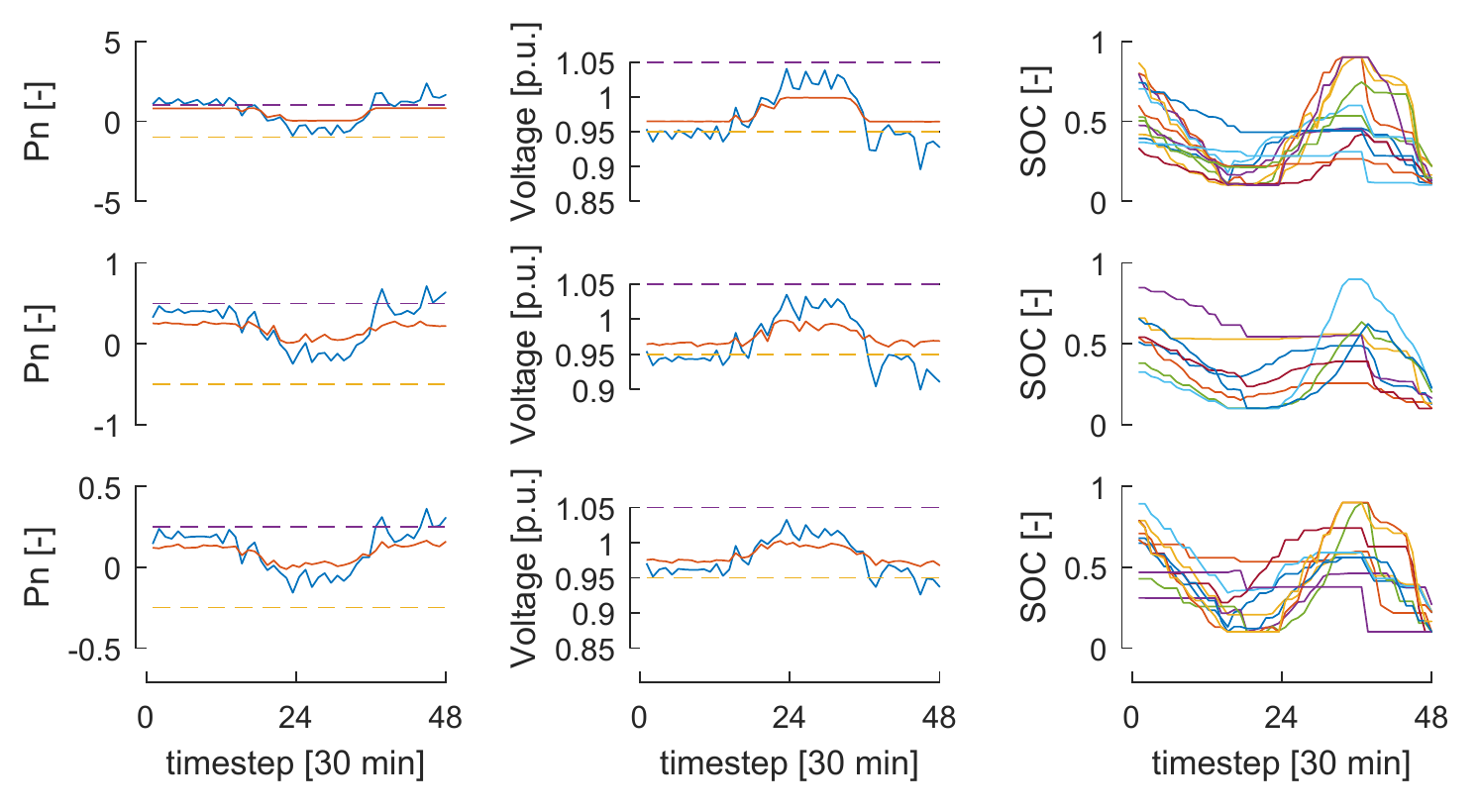}
	\caption{Example case. The considered hierarchical structure has 4 levels, and a single branching node in the first 3 levels. For the first two columns, the first row refers to the root node, second and third rows to the second and third level. First column represents the aggregated power profiles. Blue line: no battery actions. Red line: optimized power profiles. Dashed lines: power constraints. The  second column represent voltage profiles. Blue line: no battery actions. Red line optimized voltage profiles. voltages and the state of charge (SOC) of the batteries are shown. The third column represents state of charge of prosumers'batteries in the second, third and fourth levels.} 
	\label{fig:one_case}
\end{figure*}
In this section we present the results of the numerical investigation of the proposed algorithm. In particular, we simulated 500 scenarios of different hierarchical structures in order to study the algorithm performances in terms of computational time. In each scenario the prosumers coordinate their actions for the day ahead. Each agent has a random generated power profile and an electrical battery with a random starting state of charge. The battery are considered to be dynamic linear systems, cyclic and calendar aging are not considered.
For each scenario we built a random tree with at most 4 aggregator levels, which means that $l \in [2,5]$. We only consider trees in which each branching node is the parent of at most other 2 branching nodes, while the maximum number of leaf nodes per branch is 10. Only leaf nodes are considered to be flexible nodes, which means that all prosumers are located in leaf nodes, while branching nodes are aggregators. Voltage sensitivity coefficients for each level are randomly generated.
\begin{figure}[h]
	\centering
	\includegraphics[width=1\columnwidth]{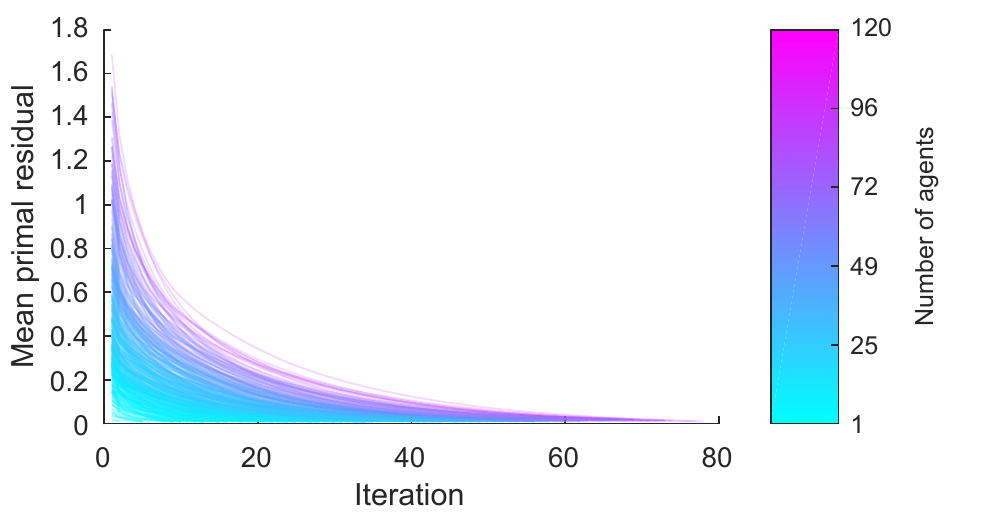}
	\caption{Mean overall primal residual for all the simulations, plotted versus the number of iterations before convergence, colored against the number of agents.}
	\label{fig:primal}
\end{figure}
With these rules, we obtain a tree with maximum number of 15 branching nodes (including the root node). The objective function of each prosumers $f_i$ is the sum of its electricity costs or revenues:
\begin{equation}
f_{i}(x_i) = \sum_{t=1}^{T} C_{i,t} 
\end{equation}
where the cost at time $t$ is defined as:
\[
C_{i,t} = 
\begin{cases}
(P_{u_i,t} + x_{i,t} )p_b, & \text{if } \quad P_{u_i,t} + x_i \ \geq \ 0\\
-(P_{u_i,t} + x_{i,t} )p_s,              & \text{otherwise}
\end{cases}
\]
where $x_i$ is the overall power from the battery, $P_{u_i}$ is the uncontrolled power of agent $i$, $p_b$ and $p_s$ are the buying and selling energy prices, respectively. Note that positive powers are considered as consumed quantities.
The system-level objective is a tracking objective with a zero power profile, which results in a quadratic peak shaving:
\begin{equation}
e(x) = \Vert S_{\emptyset} (x+P_u)  \Vert_2^2
\end{equation} 
where $P_u = [P_{u_i}] $.
The simulations are carried out using an Intel Core i7-4790K CPU @ 4.00GHz with 32 GB of RAM.

In figure \ref{fig:one_case} an example of the coordination mechanism is shown. The considered hierarchical structure has 4 levels, and a single branching node in the first three levels. Aggregated power profiles, voltages and the state of charge (SOC) of the batteries are shown. 

In figure \ref{fig:primal} the mean overall primal residual for all the simulations is shown, where the primal residual in branch $B$ is $err_B = S_B x -y_B$, is plotted versus the number of iterations before convergence, which is considered reached when $err \leq 1e-2$. The line color is related to the total number of prosumers in the related tree. As expected the number of iterations before convergence increases with the number of coordinated prosumers.

\begin{figure}[h]
	\centering
	\includegraphics[width=1\columnwidth]{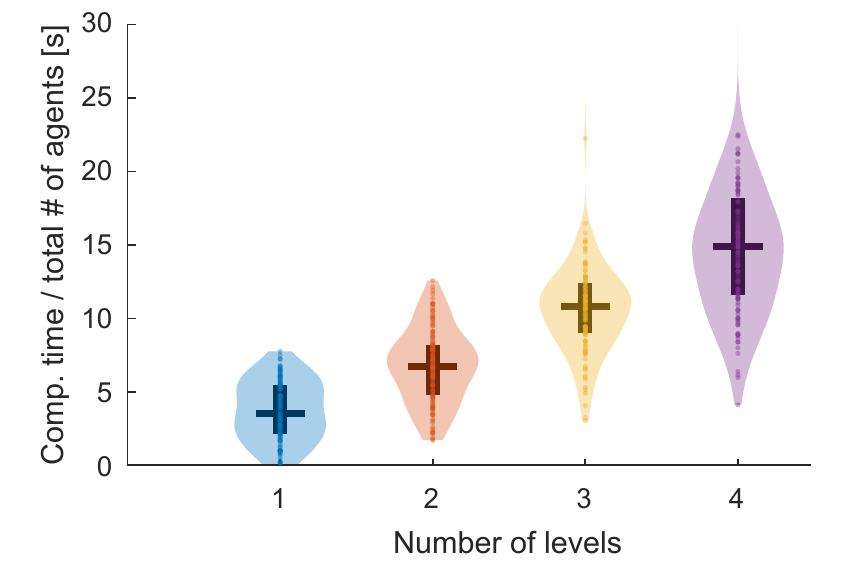}
	\caption{Estimated probability densities of the computational time divided by the total number of agents, as a function of the number of branching levels. The vertical bar is the interquartile range, the horizontal line is the median.}
	\label{fig:violin}
\end{figure}

Figure \ref{fig:violin} shows the estimated probability densities of the agent-normalized computational time, as a function of the number of branching levels of the considered tree.

\section{Conclusions}
We presented a constrained networked optimization algorithm for the coordination of prosumers, which exploits a hierarchical structure, reflecting the hierarchy of the different voltage levels of the electrical grid. Prosumers are coordinated with the help of aggregators, located at the branching nodes. The monolithic optimization problem is decomposed and parallelized using the ADMM, resulting in a forward-backward communication flow in the hierarchy.
The proposed mechanism ensures that prosumers' privacy is preserved, since communication between different levels involves only aggregated information. The numerical simulations show that the computational time normalized with the number of prosumers scales linearly with the number of levels. In future work the authors will investigate the algorithm performance using low and medium voltage test grids, by means of power flow simulations.


\section*{Acknowledgment}
The authors would like to thank Innosuisse - Swiss Innovation Agency (CH) and SCCER-FURIES - Swiss Competence Center for Energy Research - Future Swiss Electrical Infrastructure for their financial and technical support to the research work presented in this paper. This work has been sponsored by the Swiss Federal Office of Energy (Project
nr. SI/501499)



%

\bibliographystyle{IEEEtran}

\end{document}